# Potential of attosecond coherent diffractive imaging

Arjun Rana[1*], Jianhua Zhang[1,2*], Minh Pham[3], Andrew Yuan[1], Yuan Hung Lo[1,4], Huaidong Jiang[2], Stanley Osher[3] and Jianwei Miao[1†]

[1]Department of Physics & Astronomy and California NanoSystems Institute, University of California, Los Angeles, CA 90095, USA.

[2]School of Physical Science and Technology, ShanghaiTech University, Shanghai, 201210, China.

[3]Department of Mathematics, University of California, Los Angeles, CA 90095, USA.

[4]Department of Bioengineering, University of California Los Angeles, CA 90095, USA.

[†]Corresponding author. E-mail: miao@physics.ucla.edu

[*]These authors contributed equally to this work.

**Attosecond science has been transforming our understanding of electron dynamics in atoms, molecules and solids. However, to date almost all of the attoscience experiments have been based on spectroscopic measurements because attosecond pulses have intrinsically very broad spectra due to the uncertainty principle and are incompatible with conventional imaging systems. Here we report an important advance towards achieving attosecond coherent diffractive imaging. Using simulated attosecond pulses, we simultaneously reconstruct the spectrum, 17 probes and 17 spectral images of extended objects from a set of ptychographic diffraction patterns. We further confirm the principle and feasibility of this method by successfully performing a ptychographic coherent diffractive imaging experiment using a light-emitting diode with a broad spectrum. We believe this work clear the way to an unexplored domain of attosecond imaging science, which could have a far-reaching impact across different disciplines.**

The first demonstration of attosecond pulses in 2001 has opened a new window to probe electron dynamics in atoms and molecules with unprecedented time resolution [1-9]. With the continuing decrease of the temporal pulse duration and the increase of the photon energy range to the x-ray regime [10-14], the potential applications of attoscience could be even broader. However, attoscience experiments have been mostly limited to spectroscopic techniques due to the broad spectrum of the attosecond source [3-5]. For an attosecond pulse, its energy bandwidth ($\Delta E$) and temporal pulse duration ($\Delta t$) are fundamentally set by the uncertainty principle ($\Delta E \cdot \Delta t \geq \hbar/2$). For example, the recent experimental demonstration of 53-attosecond soft x-ray pulses reaches the carbon K-absorption edge (284 eV) with $\Delta E/E \approx 100\%$ [13]. Such broad spectrum pulses cause severe chromatic aberration for any lens-based imaging systems. Chromatic aberration, first discovered by Newton more than 300 years ago [15], is a failure of a lens to focus all colors to the same focal spot due to the change of the refractive index of the lens with the wavelength of light. A classical method to overcome chromatic aberration is the use of achromatic lenses [16]. But this method does not work for the full spectrum of electromagnetic radiation. For example, design and manufacture of achromatic lenses are extremely challenging in the x-ray regime, where lens design is difficult even in the monochromatic case [17]. Here we develop a more general computational method based on coherent diffractive imaging (CDI), which can not only eliminate chromatic aberration across the full spectrum of electromagnetic radiation, but also take advantage of the broad spectrum to simultaneously reconstruct the spectrum, probes and images at 17 different wavelengths.

CDI is a lensless imaging or computational microscopy method, where the diffraction patterns of an object are first measured and then directly phased to obtain high-resolution images [18]. Since the first experimental demonstration in 1999 [19], various forms of CDI such as plane-wave CDI, ptychography and Bragg CDI have been developed and applied to a broad range of samples in the physical and biological sciences using synchrotron radiation, x-ray free electron lasers, high harmonic generation, optical laser and electrons [18-31]. With advanced computational algorithms, broadband CDI has also been developed to deal with the low temporal coherence of the illumination source [32-34]. Ptychography, a powerful scanning CDI method, is particularly suitable for broadband imaging, which relies on collecting a series of diffraction patterns using a spatially confined probe to scan over an extensive object [22,31]. By partially overlapping the probe between adjacent scan positions, advanced algorithms can reconstruct both the probe and the complex exit wave of the object [23]. More recently, multimode and multiplex ptychographic methods have been developed to deal with broadband data [35-39]. Here we make a significant advance to merge CDI with attosecond science for the first time, allowing the simultaneous reconstruction of the probe, object and spectrum of attosecond pulses with $\Delta E/E \approx$ 100%.

For attosecond light with a broad spectral bandwidth, its diffraction pattern is composed of an incoherent sum of diffracted intensity from all wavelengths,

$$I_j(\vec{k}) = \int |\Psi_{\lambda,j}(\vec{k})|^2 d\lambda , \qquad (1)$$

where $I_j(\vec{k})$ is the diffraction pattern of the $j^{th}$ scan position and $\Psi_{\lambda,j}(\vec{k})$ is the diffracted wave with wavelength $\lambda$. For simplicity, we ignore the constant in front of the integral of the equation. In practice we have to discretize Eq. (1), which introduces a discretization error. In addition, noise is present in the diffraction intensity of each wavelength. By introducing a new term, $N_{\lambda,j}(\vec{k})$, to absorb the discretization error and noise, we re-write Eq. (1),

$$I_j(\vec{k}) = \sum_{\lambda=1}^{M} \left[ |\Psi_{\lambda,j,e}(\vec{k})|^2 + N_{\lambda,j}(\vec{k}) \right] = \sum_{\lambda=1}^{M} \left[ |\Psi_{\lambda,j,e}(\vec{k})|^2 + |\Psi_{\lambda,j,g}(\vec{k})|^2 \right], \qquad (2)$$

where *M* is the number of spectral images to be reconstructed, $\Psi_{\lambda,j,e}(\vec{k})$ and $\Psi_{\lambda,j,g}(\vec{k})$ represent real and ghost modes of the diffracted wave, respectively. The introduction of ghost modes is to separate the signal from the error and noise. To simultaneously reconstruct the spectrum, probes and spectral images, we develop the Spectrum, Probe and Image REconstruction (SPIRE) algorithm by incorporating the ghost mode and probe replacement constraint, whereas probe replacement identifies the best probe among all the wavelengths in each iteration and uses it to refine the probes of other wavelengths. To implement SPIRE, a set of diffraction patterns are first collected by performing a 2D scan of a confined probe relative to a sample with the probe at each position overlapped with its adjacent ones. The algorithm then iterates between real and reciprocal space (Supplemental text). In real space, the exit wave of the real and ghost modes at different wavelengths is obtained by multiplying probes by the object functions of the sample. For ghost modes, a 2D reflection transformation is applied to the coordinate of each object function. Taking the Fourier transform of the exit wave generates the diffracted wave of real and ghost modes. An updated diffracted wave is obtained by constraining it to the diffraction patterns. By applying the inverse Fourier transform, a new exit wave is created from the diffracted wave, which is used to create the next-iteration probes and object functions of real and ghost modes. To apply probe replacement, in each iteration the best probe is identified among all the wavelengths and is propagated back to construct the probes of other wavelengths. By incorporating ghost modes and probe replacement, SPIRE can simultaneously reconstruct the spectrum, probes and spectral images when the illumination source is truly broadband. The algorithm is robust as it is not sensitive to the initial input and converges after several hundreds of iterations.

To validate the SPIRE algorithm, we first performed numerical simulations using attosecond pulses with a spectrum consisting of 500 discretized modes ranging from 4.1 nm to 12.4 nm [13]. The probe was confined by a 3-μm-diameter pinhole that was placed 100 μm upstream of the sample and Fresnel propagated to the sample plane. Two samples were used in the simulations. The first is a resolution pattern, composed of 200-nm-thick aluminum structure with bar widths ranging from 30 nm to 1.2 μm. The second sample is a letter pattern with "atto" made of 200-nm-thick aluminum and "CDI" of 200-nm-thick boron. Boron was chosen because its K absorption edge (6.6 nm) is within the simulated spectrum, providing a contrast difference to assess the reconstruction quality of the spectrum and spectral images. Each sample was scanned in a randomly perturbed 2D raster grid scheme with 94% overlap between adjacent scan positions. Broadband ptychography necessitates relatively high overlap because the amount of information being recovered is larger. The resolution pattern dataset consists of 1456 scan points and the letter pattern dataset consists of 676 scan points. A flux of $1 \times 10^7$ photons per scan position was used in the simulations and Poisson noise was added to the diffraction intensity. Each diffraction pattern was collected by a detector positioned 10 cm downstream of the sample to satisfy the oversampling requirement for all wavelengths [40]. The quantum efficiency of the detector as a function of the spectrum was taken into account in each diffraction pattern. Figure 1a and b show a representative diffraction pattern from the resolution and letter pattern, respectively. The most noticeable feature of these broadband diffraction patterns is the absence of strong speckles that are presented in monochromatic diffraction patterns.

From the diffraction patterns, we used the SPIRE algorithm to reconstruct the spectrum, probes and spectral images. All reconstructions consist of two runs of 250 iterations each. In the first run, the initial guesses of the probes and spectral images were binary masks and random arrays, respectively. The second run was initialized with new random arrays of the images while retaining the reconstructed probes from the first run. In the second run, only the images were allowed to update while the probes are fixed. In all reconstructions, we chose 17 probes and 17 spectral images that span the simulated spectrum in equal wavelength intervals. The number of probes and spectral images was heuristically chosen in a manner to reconstruct the spectrum with high accuracy while also minimizing the crosstalk between adjacent images. Figure 1d-e show the reconstructed spectra (blue) of the resolution and letter pattern, which are in good agreement with the true ones (red). In comparison, a state-of-the-art broadband algorithm known as ptychographic information multiplexing (PIM) [36] failed to reconstruct the spectra in both cases (Supplemental Fig. 1). Figure 2 shows the SPIRE reconstruction of the probes and spectral images of the resolution pattern at three representative wavelengths (9.3 nm, 6.72 nm, 6.2 nm), in which the spatial resolution is increased with the decrease of the wavelength. The full reconstructions of 17 probes and 17 spectral images are shown in Supplemental Figs. 2a and 3a, respectively. SPIRE faithfully reconstructed all the probes and spectral images except the first and last image due to the low incident flux at these two wavelengths (Fig. 1d). Figure 3, Supplemental Figs. 4a and 5a, and Video 1 show the reconstructed probes and spectral images of the letter pattern. While the absorption of aluminum is relatively flat across the spectrum, the K-edge of boron at 6.6 nm causes a jump in the absorption contrast of the "CDI" letters (Fig. 3). In comparison, PIM failed to reconstruct several probes and spectral images of both the resolution and letter patterns (Supplemental Figs. 2b, 3b, 4b and 5b). To account for the spectral instability of attosecond pulses, we conducted another simulation with a shot-to-shot spectral fluctuation of 10%. With all the other parameters kept the same, SPIRE successfully reconstructed the average spectrum, probes and spectral images at 17 different wavelengths (Supplemental Figs. 6 and 7).

Next, we performed a broadband light-emitting diode (LED) experiment to validate the method. A collimated white LED was used to illuminate a test pattern with a 200 μm pinhole placed approximately 6 mm upstream of the sample to confine the probe. A CCD camera from Princeton Instruments was placed 26 cm downstream of the sample to collect diffraction patterns while fulfilling the oversampling requirement for all wavelengths of the LED[40]. A field of view approximately 600 x 600 μm was scanned using a 2D raster grid consisting of 950 points with 94% overlap between adjacent probe positions. A small random offset was applied to the scan positions to avoid gridding artifacts in the reconstructions. Three exposures of different duration were collected at each scan position and merged to improve the dynamic range of the diffraction patterns. Diffraction patterns were cropped to be square and binned by a factor of 2 in each dimension to increase the signal-to-noise ratio and reduce the computation time. After two runs of 250 iterations each, SPIRE reconstructed the spectrum, 17 probes and 17 spectral images from the diffraction patterns. The spectrum agrees with the experimental measurement using a spectrometer (Fig. 1f). Figure 4 shows three representative probes and spectral images at 662.5 nm, 550 nm and 437.5 nm. All 17 probes and images are shown in Supplemental Fig. 8. The successful reconstruction of the spectrum, probes and spectral images from the experimental diffraction patterns is consistent with the numerical simulation results and further corroborates the method.

In conclusion, we have developed a powerful algorithm (SPIRE) by introducing two new constraints - ghost modes and probe replacement. Using a simulated attosecond source and an LED with a broad spectrum, we validated the SPIRE algorithm and demonstrated the potential of achieving attosecond CDI by simultaneously reconstructing the spectrum, probes and spectral images at 17 different wavelengths. The significance of this work is threefold. First, our method is in principle applicable to any photon and electron source with a broad spectrum such as synchrotron radiation pink beams, HHG and electron sources, allowing the recovery of chemically specific contrast information without the use of monochromatic optics. Second, in this work we not only just overcome the broadband imaging obstacle, but also turn it into a strength by harnessing all the flux from a source. This would significantly reduce the data acquisition time in performing spectro-ptychography experiments, whereas the conventional method requires serial repetition of ptychographic scans as a function of the energy. Finally, this work potentially unifies two important fields - attosecond science and CDI - into a single frame. As the spectrum of the current state-of-the-art attosecond sources extends to the x-ray regime [13,14], the ability to simultaneously reconstruct the spectrum, probes and images at multiple wavelengths could find broad application ranging from visualizing attosecond electron dynamics to imaging materials and biological samples at the nanometer scale with attosecond x-ray pulses.

We thank Z. Chang, P. B. Corkum, M. M. Murnane, H. C. Kapteyn, Y. Wu, X. Ren and J. Li for stimulating discussions. This work was primarily supported by STROBE: A National Science Foundation Science & Technology Center under Grant No. DMR 1548924. We also acknowledge the support by the Office of Basic Energy Sciences of the US DOE (DE-SC0010378) and the NSF DMREF program (DMR-1437263).

**Figures and figure legends**

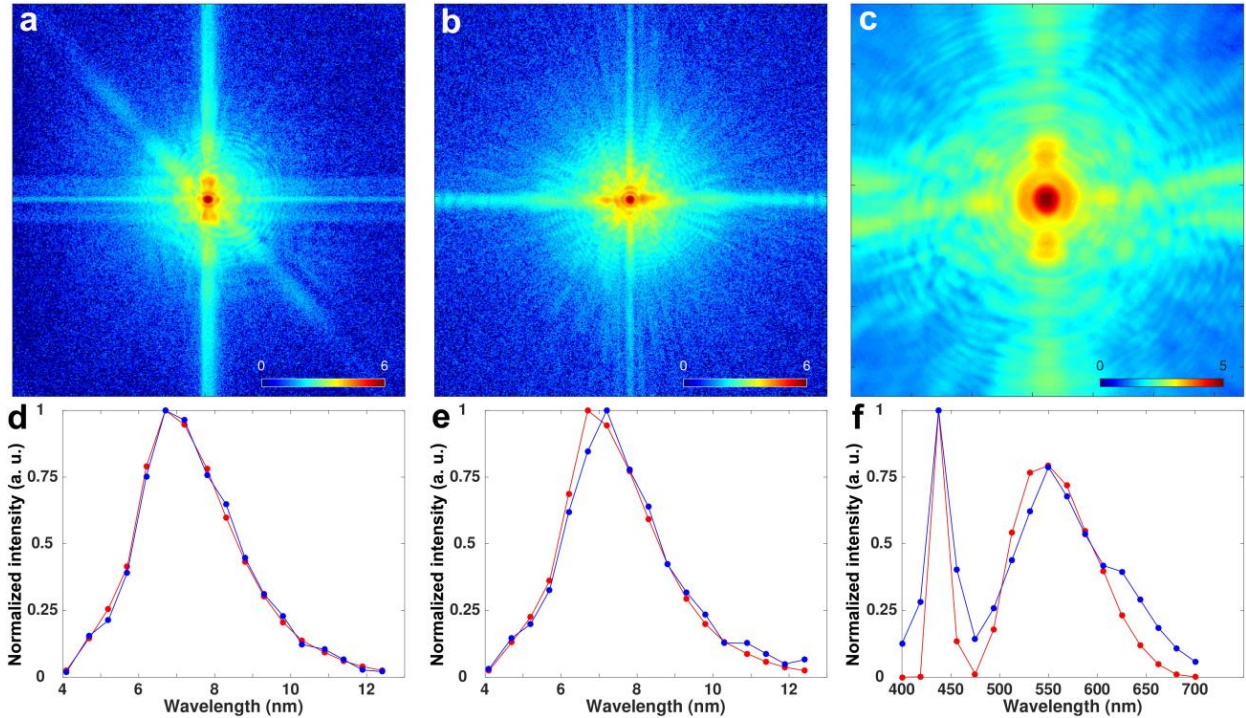

**FIG 1. Representative diffraction patterns and reconstructed spectra by SPIRE. a** and **b**, Representative diffraction patterns measured from a resolution and letter pattern, respectively, using simulated attosecond pulses. **c**, Representative diffraction pattern of a test pattern using an LED. **d**-**f**, Reconstructed spectra (in blue) of the resolution and letter pattern with simulated attosecond pulses and of the test pattern with the LED, respectively, where the true spectra are in red. The true spectrum of the LED in (**f**) was measured by a spectrometer.

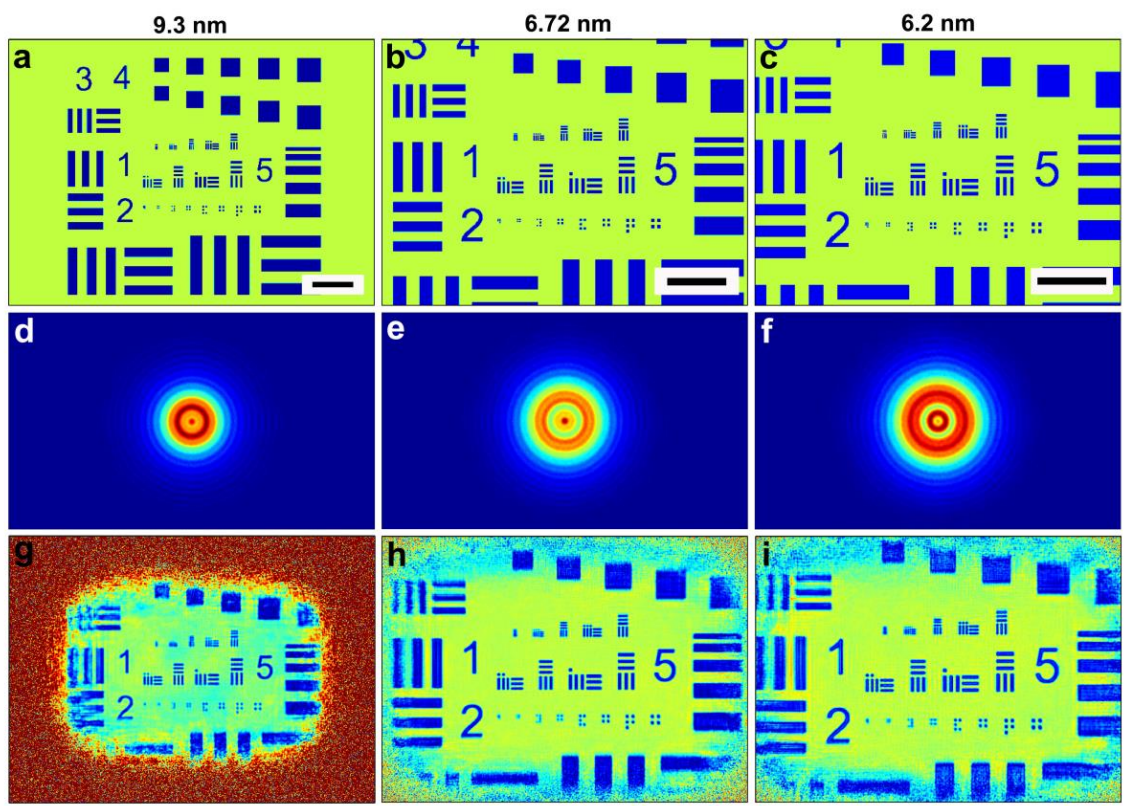

**FIG 2. Probe and spectral image reconstructions of a resolution pattern with simulated attosecond pulses. a-c**, The structure of the resolution pattern at 9.3 nm, 6.72 nm and 6.2 nm, respectively. **d-i**, Three representative probes and spectral images at different wavelengths reconstructed by SPIRE, respectively, where the spatial resolution is increased with the decrease of the wavelength. The full 17 probes and 17 spectral images are shown in Supplemental Figs. 2a and 3a, respectively. Scale bar, 2 μm.

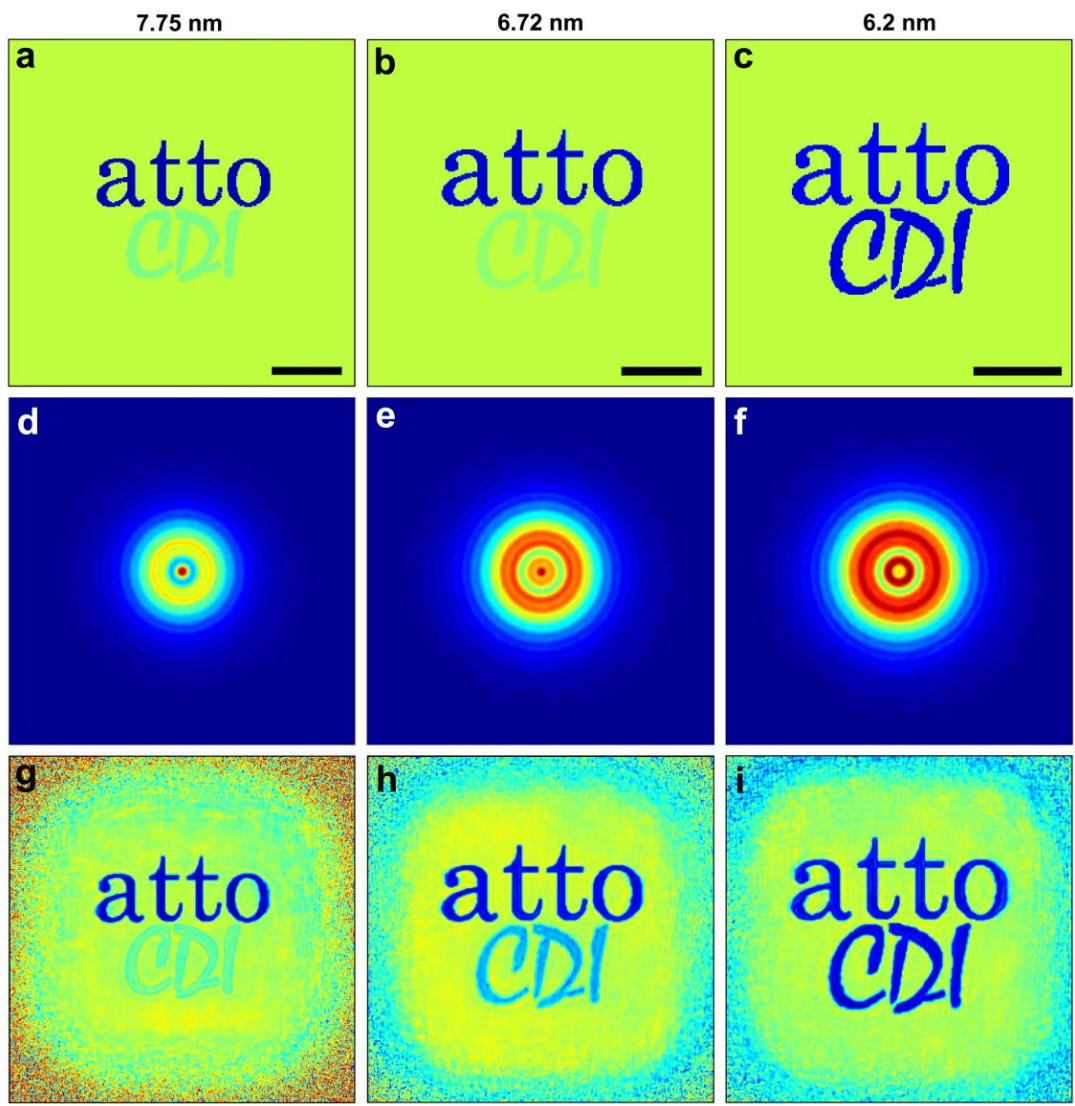

**FIG 3. Probe and spectral image reconstructions of a letter pattern with simulated attosecond pulses. a-c**, Absorption contrast images of the letter pattern at 7.75 nm, 6.72 nm and 6.2 nm, respectively, where the K-edge of boron is at 6.6 nm. **d-i**, Three representative probes and spectral images at different wavelengths reconstructed by SPIRE, respectively, where the image contrast of the "CDI" letters changes across the absorption edge. The full 17 probes and 17 spectral images are shown in Supplemental Figs. 4a and 5a, respectively. Scale bar, 2 μm.

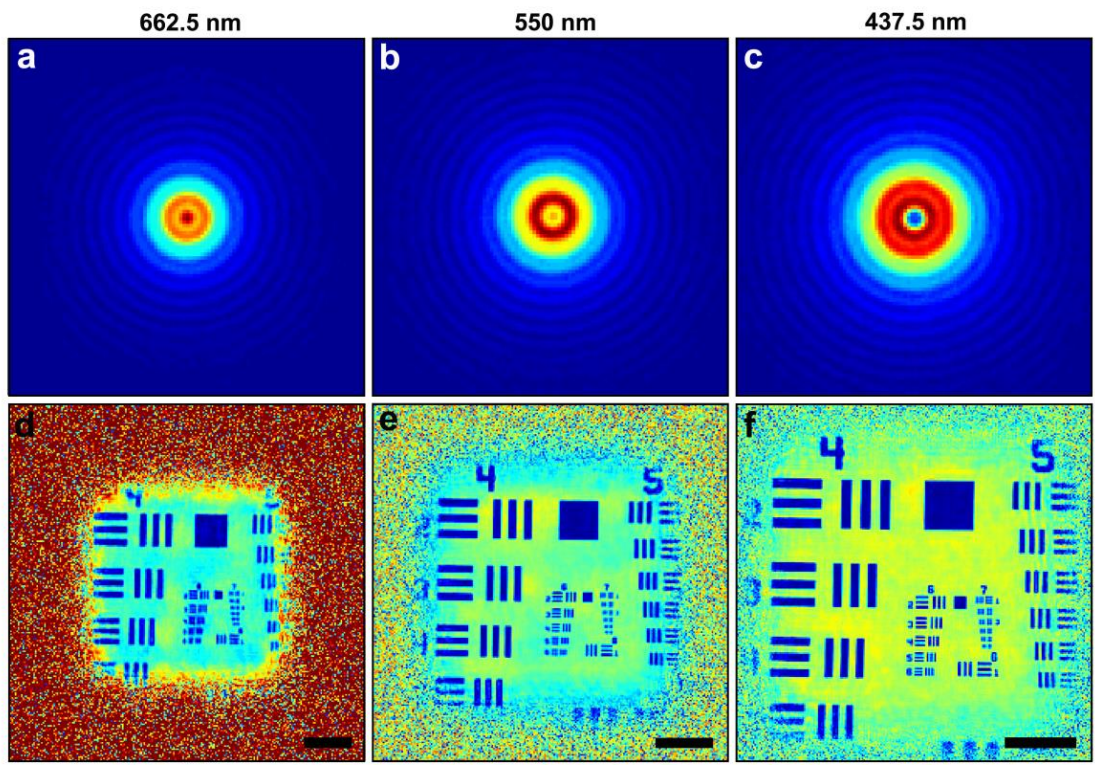

**FIG 4. Probe and spectral image reconstructions of a test pattern from a broadband LED experiment. a-f**, Three representative probes and spectral images reconstructed by SPIRE at 662.5 nm, 550 nm and 437.5 nm, respectively. The full 17 probes and 17 spectral images are shown in Supplemental Fig. 8. Scale bar, 200 μm.